\documentclass{aa}
\usepackage{txfonts}
\usepackage{graphicx}
\begin{document}
\title{High-redshift microlensing and the spatial distribution of dark matter in the form of MACHOs}
\markboth{E. Zackrisson \& T. Riehm}{High-redshift microlensing and the spatial distribution of dark matter}
\authorrunning{E. Zackrisson \& T. Riehm}
\titlerunning{High-redshift microlensing by MACHOs} 
\author{Erik Zackrisson\inst{1,2,3} \and Teresa Riehm\inst{3}}
\offprints{Erik Zackrisson, \email{ez@astro.uu.se}}
\institute{Tuorla Observatory, University of Turku, V\"ais\"al\"antie 20, FIN-21500 Piikki\"o, Finland \and Department of Astronomy and Space Physics, Box 515, S-75120 Uppsala, Sweden \and Stockholm Observatory, AlbaNova University Centre, S-10691 Stockholm, Sweden}

\date{Received 7 November 2006 / Accepted 17 August 2007}
\abstract{A substantial part of the dark matter of the Universe could be in the form of compact objects (MACHOs), detectable through gravitational microlensing effects as they pass through the line of sight to background light sources. So far, most attempts to model the effects of high-redshift microlensing by a cosmologically distributed population of MACHOs have assumed the compact objects to be randomly and uniformly distributed along the line of sight. Here, we present a more realistic model, in which the MACHOs are assumed to follow the spatial clustering of cold dark matter. Because of sightline-to-sightline variations in surface mass density, this scenario leads to substantial scatter in MACHO optical depths, which we quantify as a function of source redshift. We find that while optical depth estimates based on a uniform line-of-sight distribution are reasonable for the highest-redshift light sources, such estimates can be incorrect by a factor of $\approx 2$ for the nearby ($z\approx 0.25$) Universe. Hence, attempts to derive the cosmological density of MACHOs from microlensing observations of only a few independent sightlines can be subject to substantial uncertainties. We also apply this model to the prediction of microlensing-induced variability in quasars not subject to macrolensing, and demonstrate that relaxing the assumption of randomly and uniformly distributed MACHOs only has a modest impact on the predicted light curve amplitudes. This implies that the previously reported problems with microlensing as the dominant mechanism for the observed long-term optical variability of quasars cannot be solved by taking the large-scale clustering of dark matter into account.
}
\maketitle

\section{Introduction} 
The identification of dark matter, estimated to contribute around 90\% to the total matter content of the Universe, remains one of the most important quests of modern cosmology. To complicate the problem, dark matter appears to exist in at least two different forms: one baryonic and one non-baryonic. According to current inventories (Fukugita \cite{Fukugita}; Fukugita \& Peebles \cite{Fukugita & Peebles}), $\sim 1/3$ of the baryons ($\Omega_\mathrm{M, baryonic}\approx 0.04$) are still unaccounted for, whereas nearly all (save for a small contribution from neutrinos) of the non-baryonic matter ($\Omega_\mathrm{M,non-baryonic}=\Omega_\mathrm{M}-\Omega_\mathrm{M, baryonic}\approx 0.2$; e.g. Spergel et al. \cite{Spergel et al.}) remains elusive.   

MACHOs (Massive Astrophysical Compact Halo Objects) represent one class of dark matter candidates, which may be detected through gravitational microlensing effects as they pass through the line of sight to distant light sources (e.g. stars, supernovae, gamma-ray bursts, quasars). Although the MACHO acronym was originally invented with baryonic objects like faint stars and stellar remnants in mind (Carr \& Primack \cite{Carr & Primack}; Griest \cite{Griest}), several non-baryonic dark matter candidates\footnote{In this category, one usually also includes candidates that are formally baryonic (like quark nuggets), but which evade the current constraints on the cosmic baryon density and are expected to behave like cold dark matter.} can also manifest themselves in this way -- e.g. axion aggregates (Membrado \cite{Membrado}), mirror matter objects (Mohapatra \& Teplitz \cite{Mohapatra & Teplitz}), primordial black holes (Green \cite{Green}), quark nuggets (Chandra \& Goyal \cite{Chandra & Goyal}), preon stars (Hansson \& Sandin \cite{Hansson & Sandin}) and scalar dark matter miniclusters (Zurek et al. \cite{Zurek et al.}). Hence, MACHOs could in principle make up a substantial fraction of the dark matter of the Universe.

Potential MACHOs have indeed been detected through microlensing effects towards stars in the LMC (Alcock et al. \cite{Alcock et al.}), M31 (Calchi Novati et al.  \cite{Calchi Novati et al.}) and towards multiply-imaged quasars (e.g. Pelt et al. \cite{Pelt et al.}; Koopmans \& de Bruyn \cite{Koopmans & de Bruyn}; Colley \& Schild \cite{Colley & Schild}). Claims of MACHO microlensing of gamma-ray bursts (Garnavich et al. \cite{Garnavich et al.}; Baltz \& Hui \cite{Baltz & Hui}), and quasars which are not multiply-imaged (e.g. Hawkins \cite{Hawkins96}; Zakharov et al. \cite{Zakharov et al.}) have also been made, but in these cases, the interpretation is more ambiguous, since intrinsic variability cannot easily be ruled out. 

For the detections of MACHOs towards the LMC, alternative explanations like self-lensing (i.e. microlensing by normal stars inside the target galaxy, rather than MACHOs in the dark halo of the Milky Way), variable stars or background supernovae have been suggested (e.g. Evans \& Belokurov \cite{Evans & Belokurov}). It is also possible that blending effects introduced by the use of faint source stars may have affected the results in some undesired way (Tisserand et al. \cite{Tisserand et al.}). While these explanations cannot account for the microlensing events seen towards multiply-imaged quasars, the MACHO interpretation of some of the latter observations has nonetheless been questioned, as the poorly understood structure of these light sources make it very difficult to robustly determine the nature of the microlenses in this lensing situation (e.g. Wyithe \& Loeb \cite{Wyithe & Loeb}; Lewis \& Gil-Merino \cite{Lewis & Gil-Merino}). 

Although the local microlensing surveys impose strong upper limits on the dark matter contribution from MACHOs in the $10^{-7} \ M_\odot$--$10^1 \ M_\odot$  mass range (e.g. Lasserre et al. \cite{Lasserre et al.}; Tisserand et al. \cite{Tisserand et al.}), these constraints apply only to classes of MACHOs to which these surveys are sensitive. In fact, local microlensing observations are only able to detect the densest, most compact MACHOs at a given mass (e.g. Walker \& Wardle \cite{Walker & Wardle}; Zackrisson \cite{Zackrisson}; Zurek et al. \cite{Zurek et al.}), whereas more diffuse objects will give rise to detectable microlensing effects only at higher redshifts. 

Quasar emission-line and variability statistics also impose a number of important constraints on the cosmological density of MACHOs (Schneider \cite{Schneider}; Dalcanton et al. \cite{Dalcanton et al.}; Wiegert \cite{Wiegert}), but these are sensitive to assumptions concerning the typical size of the optical continuum-emitting region in quasars  and can in principle be sidestepped if this parameter is taken to be sufficiently large (Zackrisson \& Bergvall \cite{Zackrisson & Bergvall}; Zackrisson et al. \cite{Zackrisson et al. b}). Disturbingly, the recent analysis by Pooley et al. (\cite{Pooley et al. a,Pooley et al. b}) of combined X-ray and optical data on multiply-imaged quasars suggests that their optical continuum-emitting regions may be much larger than is usually assumed.

Thus, the existence and relative importance of MACHOs is still a matter of debate. To remedy this unsatisfactory situation, a large number of new microlensing surveys at both low and high redshift are currently being planned or are already underway. While the SuperMACHO project is continuing to monitor the LMC, the Angstrom, MEGA, POINT-AGAPE and WeCapp collaborations are targeting M31. Upcoming high-resolution facilities like the SIM PlanetQuest satellite also show great promise for the detection of astrometric microlensing effects, which will help to constrain the properties of the local MACHOs (e.g. Paczynski \cite{Paczynski}, Han \& Kim \cite{Han & Kim}, Rahvar \& Ghassemi \cite{Rahvar & Ghassemi}).

To assess the possibility of more diffuse MACHOs and MACHOs on larger (galaxy-cluster) scales, microlensing monitoring of M87 in the Virgo cluster (Baltz et al. \cite{Baltz et al.}) has started, and several other high-redshift MACHO searches based on cluster-quasar microlensing (Tadros et al. \cite{Tadros et al.}) and cluster-cluster microlensing (Totani \cite{Totani}) have also begun. With future supernova satellites like SNAP, JEDI or Destiny, it should also be possible to impose strong constraints on MACHOs from effects on the distribution of type Ia supernova peak luminosities (e.g. Minty et al. \cite{Minty et al.}; Metcalf \& Silk \cite{Metcalf & Silk}) and by the temporal light curve fluctuations (e.g. Bagherpour et al. \cite{Bagherpour et al.}; Dobler \& Keeton \cite{Dobler & Keeton}) that microlensing by foreground objects is expected to produce. Microlensing by MACHOs can also leave an imprint in the afterglows of gamma-ray bursts (e.g. Wyithe \& Turner \cite{Wyithe & Turner b}; Baltz \& Hui \cite{Baltz & Hui}). 

The interpretation of microlensing effects at cosmological distances is much more challenging than for local light sources. Due to the higher optical depth, there is a non-negligible probability that several microlenses may simultaneously contribute to the magnification. In the case of multiply-imaged light sources, it is customary to adopt the thin lens approximation, as most of the microlensing can be assumed to take place in the dark halo responsible for the macrolensing effect. For high-redshift light sources (i.e. quasars, supernovae, gamma-ray bursts) which are not multiply-imaged, the situation becomes more complicated, as compact objects at vastly different distances along the line of sight may contribute to the microlensing. Most studies of the latter situation (e.g. Rauch \cite{Rauch}; Schneider \cite{Schneider}; Loeb \& Perna \cite{Loeb & Perna}; Minty et al. \cite{Minty et al.}; Zackrisson \& Bergvall \cite{Zackrisson & Bergvall}; Zakharov et al. \cite{Zakharov et al.}; Baltz \& Hui \cite{Baltz & Hui}) adopt the Press \& Gunn (\cite{Press & Gunn}) approximation, in which the microlenses are assumed to be randomly and uniformly distributed with constant comoving density along the line of sight. Due to the strong clustering of matter, this approximation may not be entirely suitable and may give rise to misleading results (Wyithe \& Turner \cite{Wyithe & Turner a}). In this paper, we relax this approximation in favour of a model where the compact objects instead follow the spatial distribution of cold dark matter, thereby clustering into cosmologically distributed halos and subhalos, as would be expected for non-baryonic MACHOs. We demonstrate that this gives rise to substantial sightline-to-sightline scatter in microlensing optical depths, and quantify this scatter as a function of source redshift. By applying the model to the microlensing contribution to the long-term optical variability of quasars which are not multiply-imaged, we also show that relaxing the Press \& Gunn approximation only has a modest effect on the prediction of light curve amplitudes. Hence the conclusion reached by Zackrisson et al.  (\cite{Zackrisson et al. a}), who argued that the long-term optical variability of quasars could not primarily be caused by microlensing -- contrary to claims by Hawkins (e.g. \cite{Hawkins96}, \cite{Hawkins00}, \cite{Hawkins01}, \cite{Hawkins02}) -- appears to be robust to the line-of-sight variations in matter content predicted in the $\Lambda$CDM cosmology.

The matter distribution model and its numerical implementation are described in Sect 2. In Sect. 3, we apply this model to the calculation of the microlensing optical depth for high-redshift light sources, and explore the impact of the various components of the model. In Sect. 4, we couple the matter distribution model to the microlensing code by Zackrisson \& Bergvall (\cite{Zackrisson & Bergvall}), and apply it to problem of predicting the microlensing contribution to the long-term optical variability of quasars. Sect. 5 discusses the simplifications adopted in this work and a number of ideas for further refinement. Our findings are summarized in Sect. 6.

\section{The distribution of dark matter along the line of sight to light sources at cosmological distances}
The matter distribution model used here is based on the $\Lambda$CDM scenario, where we assume $\Omega_\mathrm{M}=0.3$, $\Omega_\mathrm{bar}=0.04$ $\Omega_\Lambda=0.7$, $\sigma_8=0.9$ and $H_0=100h$ km s$^{-1}$ Mpc$^{-1}$ with $h=0.72$. The overall results, however, are not sensitive to minor adjustments of these parameters, like the ones suggested by the 3-year WMAP data (Spergel et al. \cite{Spergel et al.}). 

\subsection{MACHOs}
As the concept of MACHOs (Carr \& Primack \cite{Carr & Primack}; Griest \cite{Griest}) was introduced before the emergence of the $\Lambda$CDM model, and before non-baryonic matter had reached its current stature, it is not obvious how these objects should be defined, or how they fit into the hierarchical picture of modern cosmology. Should only point-mass objects, or objects with densities above a certain absolute threshold be considered MACHOs? Do MACHOs populate all halos, or just halos of a certain mass (e.g. galaxy-mass halos)? What about the low-density material that cannot easily be identified as belonging to any halo? 

Throughout this paper, we use the MACHO acronym to describe any kind of dark matter objects sufficiently compact to cause microlensing effects at cosmological distances, even though this definition converts into a distance-dependent density threshold (e.g. Zackrisson \cite{Zackrisson}). We furthermore assume that MACHOs follow the spatial distribution of CDM, even when some fraction of the CDM is assumed to be smoothly distributed, i.e. either not associated with halos, or associated with halos below the mass resolution limit of our numerical model. 
 
\subsection{Dark halos}
We assume that a redshift-dependent mass fraction $f_\mathrm{halo}(z)$ of all matter is associated with dark matter halos of masses $M$ in the interval $M_\mathrm{min}\leq M \leq M_\mathrm{max}$, whereas a fraction $f_\mathrm{smooth}(z)=1-f_\mathrm{halo}(z)$ is smoothly distributed. We furthermore assume that MACHOs constitute a scale- and redshift-independent mass fraction $f_\mathrm{MACHO}= \Omega_\mathrm{MACHO}/\Omega_\mathrm{M}$ of the matter in both of these components, as the smooth component may indeed contain halos at masses below the adopted resolution limit. From now on, we reserve the term halo for objects that are not located inside other halos, and the term subhalos for those that are. For the mass function and spatial distribution of the halos, two options are implemented. Either a halo catalogue from an N-body simulation is used, providing both the masses and positions of the halos, or the halos are assumed to be randomly distributed throughout the simulated volume, obeying the redshift-dependent mass function given by Sheth \& Tormen (\cite{Sheth & Tormen}):
\begin{equation}
\frac{dn(M,z)}{d\log M} = \frac{\rho(z)}{M}\frac{d\ln \sigma^{-1}}{d\log M}A\sqrt{\frac{2a}{\pi}}\left[1+(\frac{\sigma^2}{a\delta^2})^p \right]\frac{\delta}{\sigma}\exp\left[ \frac{-a\delta^2}{2\sigma^2} \right]
\label{haloMFeq}
\end{equation} 
where $n(M,z)$ is the number density of halos with masses below $M$ at redshift $z$, $\rho(z)$ is the mean matter density of the Universe, $A=0.3222$, $p=0.3$ and $a=0.707$.

Each dark matter halo is assumed to have a smooth, spherical component obeying the Navarro et al. (\cite{NFW96,NFW97}; hereafter NFW) density profile, given by:
\begin{equation}
\rho(r)=\frac{\rho_\mathrm{i}}{(r/r_\mathrm{S})(1+r/r_\mathrm{S})^2},
\label{NFWeq}
\end{equation} 
where $r_\mathrm{S}$ is the characteristic scale radius of the halo and $\rho_\mathrm{i}$ is related to the density of the Universe at the time of collapse.  The halos are truncated at the virial radius $r_\mathrm{vir}(z)$, defined as the radius at which mean enclosed density equals $\Delta_{\mathrm{vir},z}$ times the critical density $\rho_{\mathrm{c},z}$ of the Universe at redshift $z$:
\begin{equation}
r_\mathrm{vir}(z)=\left( \frac{3M}{4\pi\ \Delta_{\mathrm{vir}, z}\ \rho_{\mathrm{c},z}}\right)^{1/3},
\label{rvireq}
\end{equation}
where $\Delta_{\mathrm{vir},z}$ can be approximated by (Bryan \& Norman \cite{Bryan & Norman}):
\begin{equation}
\Delta_{\mathrm{vir},z}\approx 18\pi^2+82x-39x^2,
\label{Deltaeq}
\end{equation}
with $x=\Omega_\mathrm{M}(z)-1$.\\

The virial radius is linked to $r_\mathrm{S}$ by the concentration parameter $c=r_\mathrm{vir}/r_\mathrm{S}$, with a median value of (Bullock et al. \cite{Bullock et al.}):
\begin{equation}
\bar{c}=\frac{9}{1+z}\left( \frac{Mh}{1.5\times 10^{13}\ M_\odot}  \right)^{-0.13}
\label{cmedianeq}
\end{equation}
and a scatter given by the log-normal distribution:
\begin{equation}
p(c)=\frac{1}{\sqrt{2\pi}\ c\ \sigma_{\ln c}}\exp \left[-\frac{(\ln c - \ln \bar{c})^2}{2\sigma^2_{\ln c}}\right],
\label{cdispeq}
\end{equation}
where the standard deviation is $\sigma_{\ln c}=0.3$ (Wechsler et al. \cite{Wechsler et al.}).

Given $c$, the $\rho_\mathrm{i}$ parameter can be computed, using:
\begin{equation}
\rho_\mathrm{i}=\frac{\Delta_{\mathrm{vir},z}\ \rho_{\mathrm{c},z}}{3}\frac{c^3}{\ln (1+c)-c/(1+c)}.
\label{rhoieq}
\end{equation}

\subsection{Dark halo substructure}
Within each dark matter halo, we assume that a fraction $f_\mathrm{subhalo}(z,M)$ of the halo mass $M$ is in the form of subhalos. To ensure mass conservation in the parent halo, the density $\rho(r)$ of its smooth NFW component is decreased throughout the halo by a factor $(1-f_\mathrm{subhalo})$. For the subhalo mass distribution, we adopt the semianalytical model developed by van den Bosch et al. (\cite{van den Bosch et al.}), in which the average mass fraction in subhalos for a parent halo of mass $M$ is: 
\begin{equation}
\log f_\mathrm{subhalo}=\sqrt{0.4\log (M/M_\star)+5}-2.74,
\label{subhalofraceq}
\end{equation}
Here, $M_\star$ represents the redshift-dependent characteristic mass scale defined by $\sigma(M_\star,z)=\delta_\mathrm{c}(z)$, where $\sigma^2(M_\star,z)$ is the mass variance of the smoothed density field (which is related to the CDM power spectrum; see e.g. Eisenstein \& Hu \cite{Eisenstein & Hu}) and $\delta_\mathrm{c}$ is the critical threshold for spherical collapse (e.g. Navarro et al. \cite{NFW97}).  

In the van den Bosch model, the subhalo mass function for a parent halo of mass $M$ is a Schechter function of the form:
\begin{equation}
\frac{\mathrm{d}N}{\mathrm{d}\ln \Psi}=\frac{\gamma}{\beta\Gamma(1-\alpha)} \left(\frac{\Psi}{\beta}\right)^{-\alpha} \exp\left( -\frac{\Psi}{\beta} \right),
\label{subhaloMFeq}
\end{equation}
where $\Psi=M_\mathrm{sub}/M$, $\beta=0.13$, $\alpha=0.996-0.028\log(M/M_\star)$. Here, $\gamma$ is given by:
\begin{equation}
\gamma=\frac{f_\mathrm{subhalo}}{P(1-\alpha,1/\beta) - P(1-\alpha, 10^{-4}/\beta)},
\end{equation}
and $P$ represents the incomplete Gamma function. 
The subhalo mass function (\ref{subhaloMFeq}) is assumed to be valid throughout the mass range $10^{-4}\leq \Psi \leq f_\mathrm{subhalo}$. 

Due to the strong tidal stripping experienced by subhalos and the very high resolution required by N-body simulations to realistically treat this effect, the most appropriate form for the typical density profile of a subhalos is somewhat unclear at the moment. In the gravitational lensing literature, subhalos are usually treated as either NFW spheres (e.g. Metcalf \cite{Metcalf}), singular isothermal spheres (e.g. Keeton \cite{Keeton}) or pseudo-Jaffe spheres (e.g. Dalal \& Kochanek \cite{Dalal & Kochanek}). For simplicity, we assume here that the NFW profile adequately describes both halos and subhalos. As we will argue that subhalos represent one of the least important ingredients in the matter distribution model, this choice is not critical.

Once the number and masses of the subhalos inhabiting a given parent halo have been determined using (\ref{subhalofraceq}) and (\ref{subhaloMFeq}), these objects are distributed within the parent halo with a radial distribution of the type suggested by Gao et al. (\cite{Gao et al.}):
\begin{equation}
\frac{n(<r)}{N}=\frac{(1+ac)r^\beta}{1+acr^\alpha},
\label{subhalodisteq}
\end{equation}
where $r$ is the distance to the parent halo centre in units of $r_\mathrm{vir}$, $n(<r)$ is the number of subhalos within $r$, $N$ is the total number of subhalos within $r_\mathrm{vir}$, $c$ is the concentration parameter of the parent halo, $a=0.244$, $\alpha=2$ and $\beta=2.75$. Only one generation of subhalos is assumed, i.e. the possibility of subhalos within subhalos (e.g. Shaw et al. \cite{Shaw et al.}) is neglected.

\subsection{Computational method}
To predict the matter distribution statistics for an ensemble of independent sightlines to the high-redshift Universe, a large number of sightline simulations are generated in a Monte Carlo fashion.  To model the spatial distribution of dark matter between the observer and source, this volume is divided into $i$ lens planes, equally spaced in redshift. For each lens plane, a cylindric volume with radius $R_\mathrm{cyl}$ is considered, which is taken to be large enough to encompass the relevant features of the large-scale structure of the Universe. In practice, we take this cylinder to be equal to the virial radius of the most massive halo in the local Universe: $R_\mathrm{cyl}=R_{\mathrm{vir,} z=0}(M_\mathrm{max})$. Numerical tests indicate that choosing a simulation radius larger than this does not affect the statistical results. The total mass associated with each such cylinder is $i$ taken to be:
\begin{equation}
M_{\mathrm{cyl, }i} = \Omega_\mathrm{M}(1+z_i)^3 \rho_\mathrm{c}\pi R_\mathrm{cyl}^2 \Delta D^\mathrm{c}_i
\end{equation}
where $\rho_\mathrm{c}$ is the critical density of the Universe and $\Delta D^\mathrm{c}_i$ is the light-travel time distance between the lens planes $i-i$ and $i$. In a flat Friedmann-Robertson-Walker Universe ($\Omega_M + \Omega_\Lambda=1$), $\Delta D^\mathrm{c}_i$ is given by:
\begin{equation}  
\Delta D^\mathrm{c}_i=\frac{c}{H_0}\int_{z_{i-1}}^{z_{i}}\frac{\mathrm{d}z}{(1+z)\sqrt{\Omega_\mathrm{M}(1+z)+\Omega_\Lambda}}.
\end{equation}
For each lens plane $i$, a fraction $f_{\mathrm{halo},i}$ of the mass $M_{\mathrm{cyl, }i}$ is assumed to be locked up in dark matter halos, whereas a fraction  $f_{\mathrm{smooth}, i}=1-f_{\mathrm{halo}, i}$ is taken to be smoothly distributed throughout the cylinder segment $i$. 

In scenarios involving halos with random spatial distribution, each cylindrical volume $i$ is populated with halos by random sampling of the dark halo mass function described in Sect. 2.2 until the total halo mass reaches $f_{\mathrm{halo}, i}M_{\mathrm{cyl, }i}$.  

In scenarios involving spatially clustered halos, each cylindrical volume $i$ is inserted with random impact parameter into the N-body simulation cube closest in redshift, and the properties (masses and coordinates) of the halos that happen to be located within that volume extracted. In cases where the N-body cube turns out to be smaller than the length $D^\mathrm{c}_i$ of the cylinder, the part of the cylinder falling outside the cube is inserted into a new cube with a new random impact parameter. This procedure is repeated until every cylinder segment is completely located inside an N-body simulation cube. 

Once the masses $M$ and positions of each halo inside each cylinder segment have been determined, a fraction $f_\mathrm{subhalo} (M,z)$ of each halo mass is redistributed in the form of subhalos through random sampling of the subhalo mass function and radial distribution described in Sect. 2.3. 

Finally, we integrate the mass density along an infinitesimal beam through each cylinder centre, to derive the surface mass density of matter $\Sigma_i$ in that plane compared to the cosmic average $\bar{\Sigma_i}$ at that redshift $z_i$:
\begin{equation}
f_{\Sigma, i}=\frac{\Sigma_i}{\bar{\Sigma_i}}=\frac{\Sigma_i}{\Omega_\mathrm{M}(1+z_i)^3 \rho_\mathrm{c}\Delta D^\mathrm{c}_i}.  
\label{massfrac}
\end{equation}
The stochastic variation of this quantity along each sightline and between different sightlines is then used to predict sightline-to-sightline scatter in microlensing properties.  

\subsection{Matter distribution scenarios}
To assess the importance of the various ingredients of the mass distribution model, five different levels (A--F) of refinement are tested. 

\begin{figure*}[t]
\resizebox{\hsize}{!}{\includegraphics{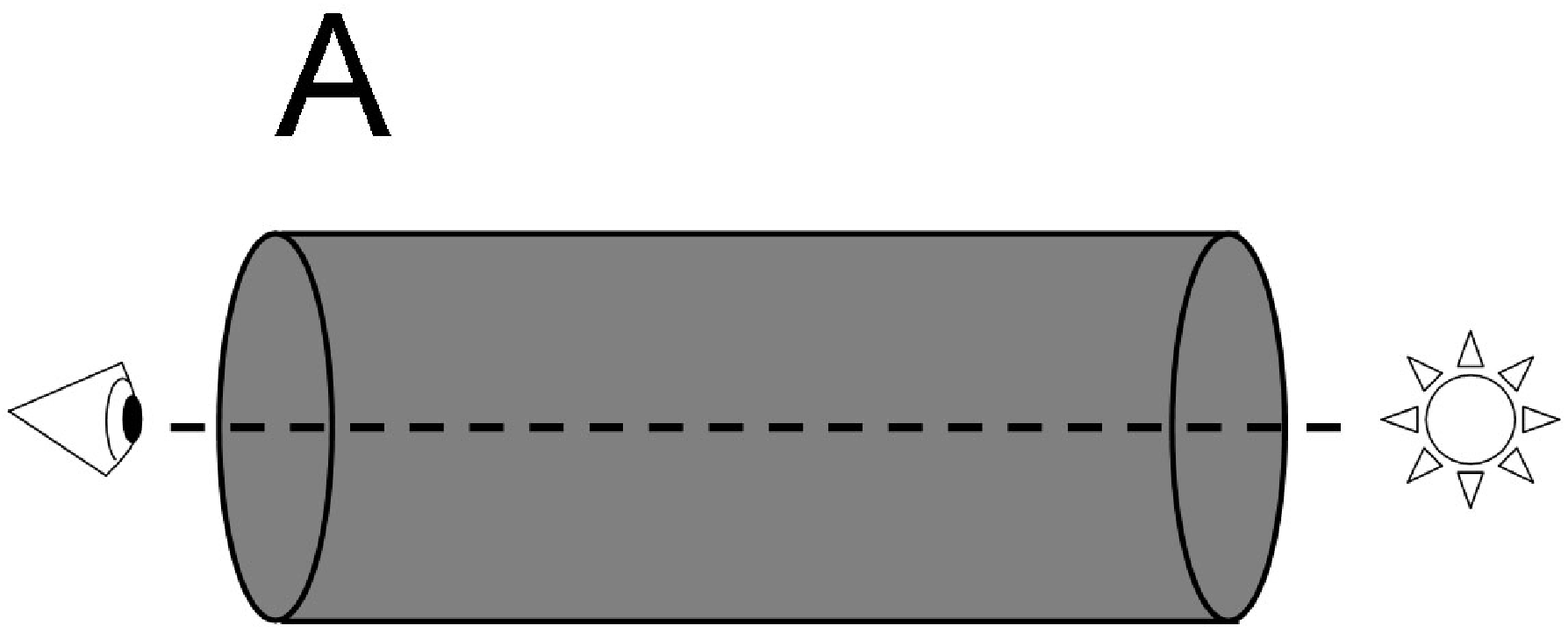}\includegraphics{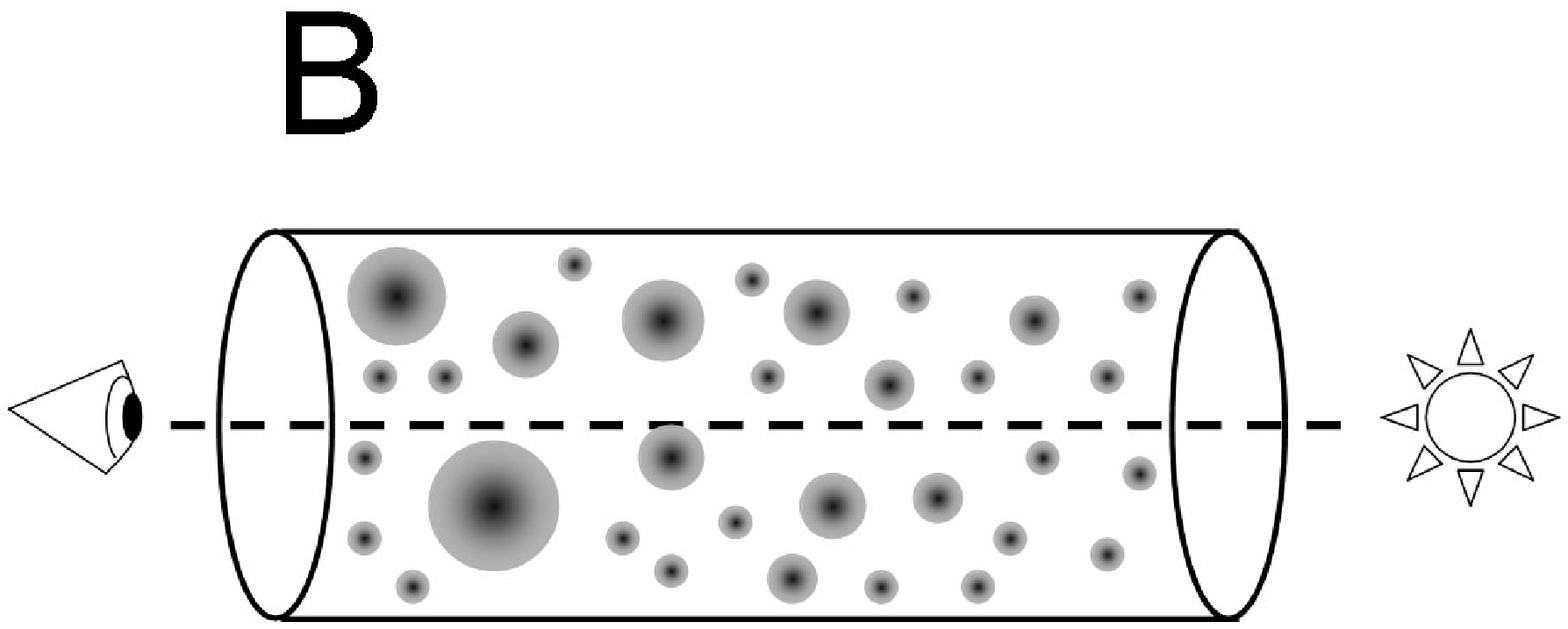}\includegraphics{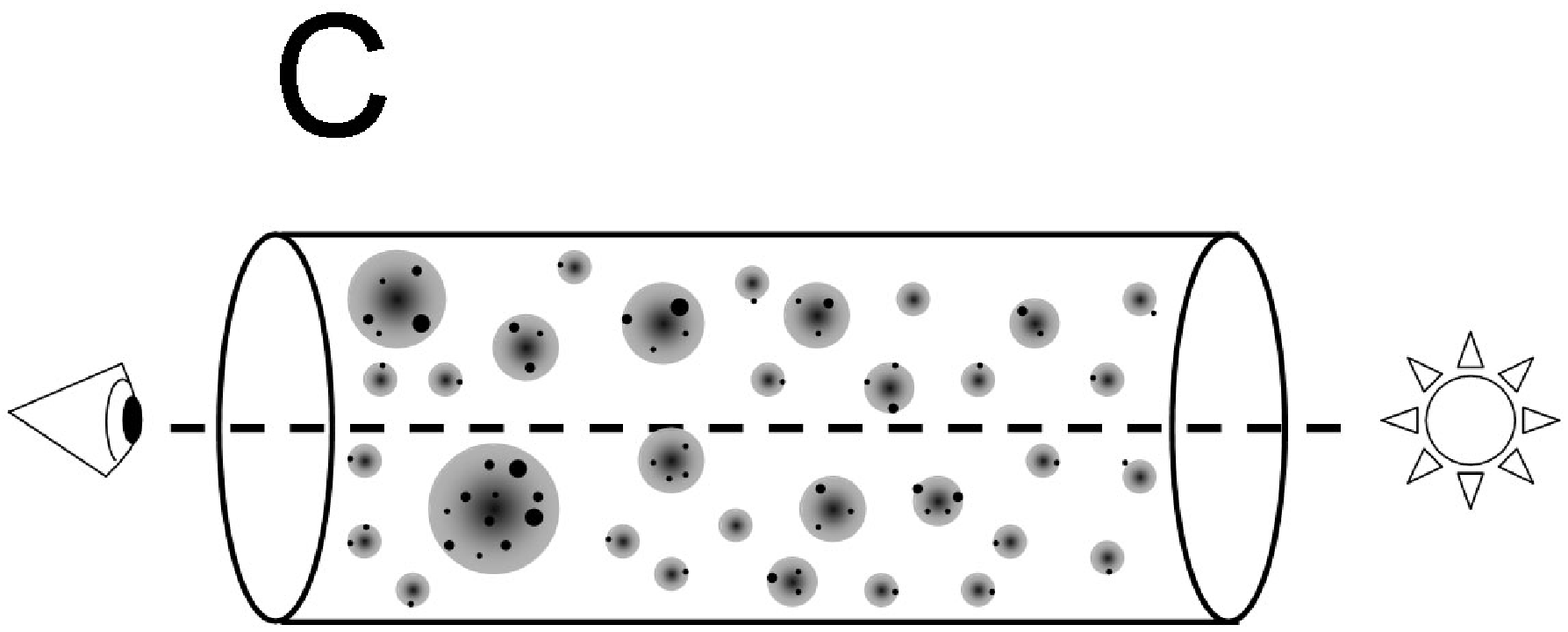}}
\resizebox{\hsize}{!}{\includegraphics{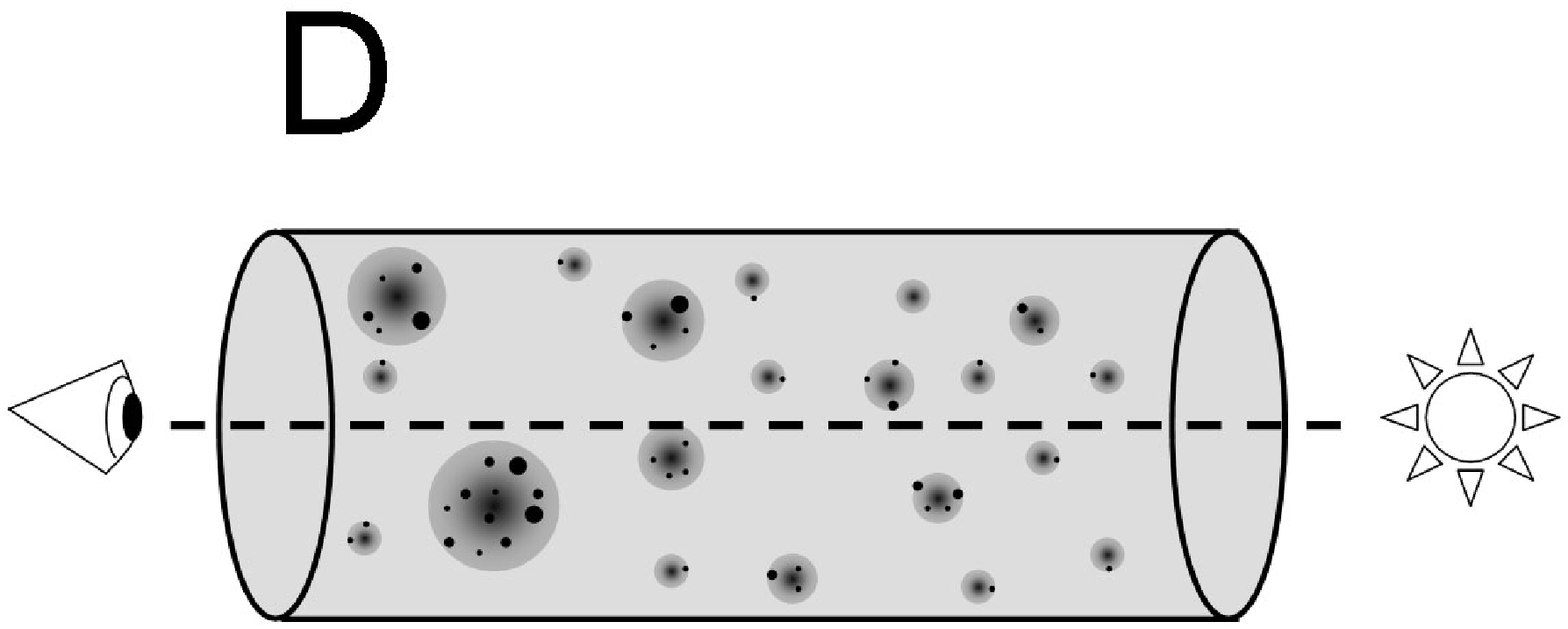}\includegraphics{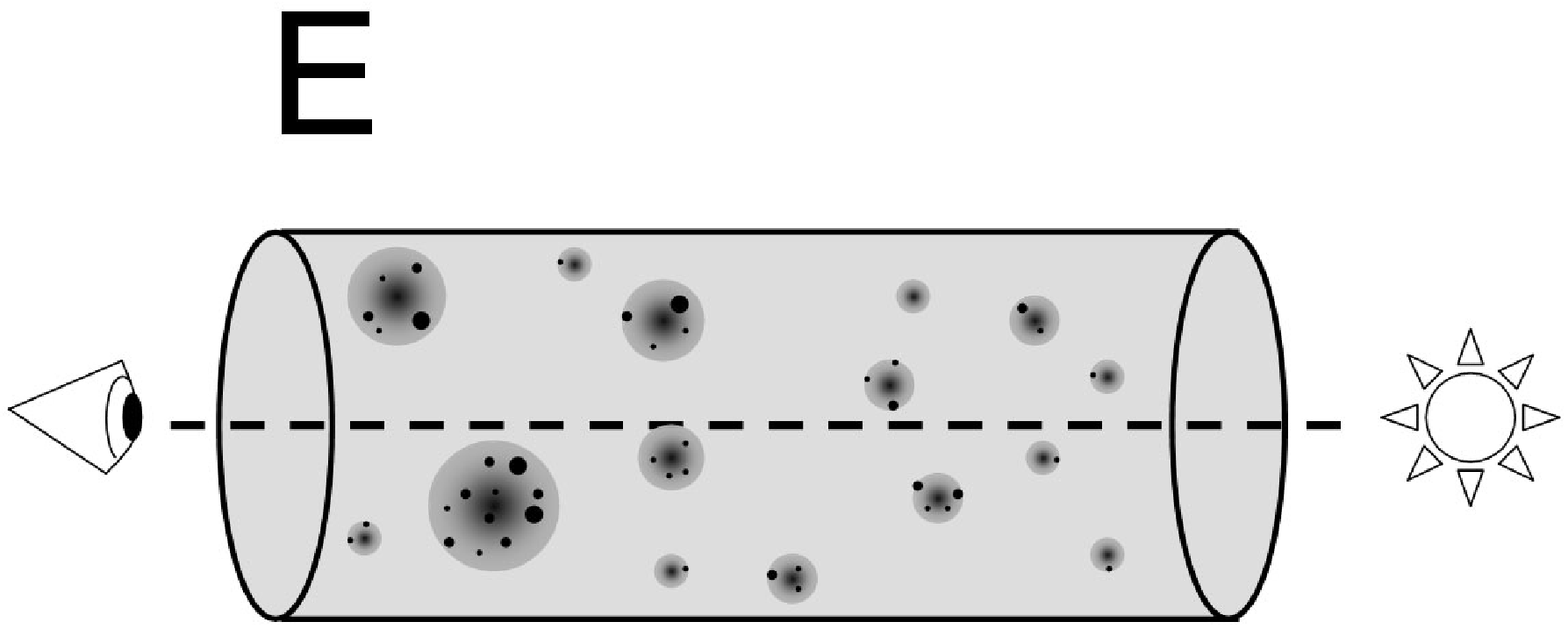}\includegraphics{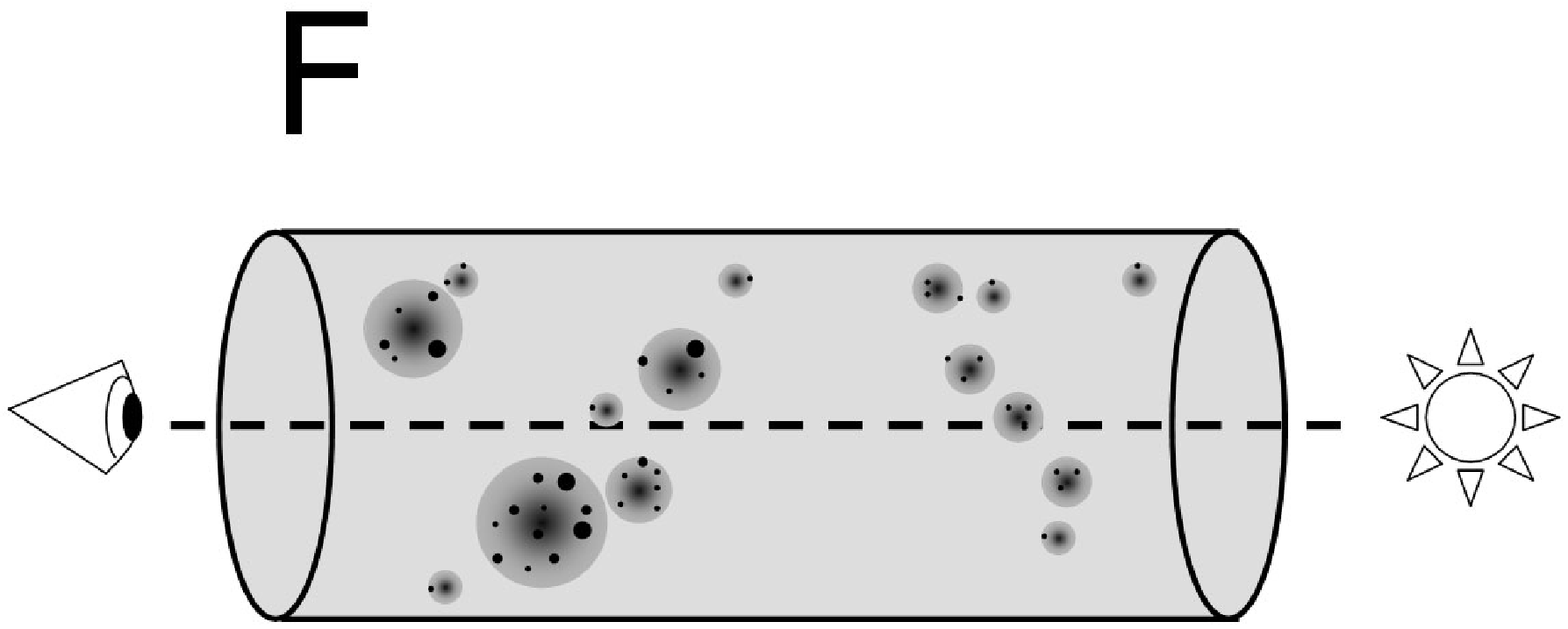}}
\caption[]{Schematic illustrations of matter distribution scenarios A--F. Here, the cylinder represents the cosmological volume between the observer and source. In scenario A, all the matter is smoothly distributed along the sightline, in accordance with the Press \& Gunn (\cite{Press & Gunn}) approximation. In scenario B, all the matter is instead assumed to be locked up inside halos with random spatial distribution. The effects of hierarchical halo clustering is here illustrated as evolution in the typical halo sizes between the source (high redshift; only low-mass halos) and observer (low redshift; both low- and high-mass halos). In C, substructure has been added to each halo. In D, a component of smoothly distributed matter (light gray) has furthermore been included to represent the matter that does not seem to be associated with halos within the mass range considered. Scenario E is similar to D, except that a halo mass catalogue from an N-body simulation has been used. Because of the resolution limit of this simulation, the number of small-mass halos has decreased. Scenario F takes into account the spatial clustering of halos predicted by N-body simulations, but is otherwise identical to scenario E.}
\label{matterdistfig}
\end{figure*}

\begin{itemize}
\item {\bf A}. All matter is assumed to be randomly distributed with constant comoving density along the sightline, i.e. according to the Press \& Gunn (\cite{Press & Gunn}) approximation.
 
\item {\bf B}. All matter is assumed to be in the form of smooth NFW halos ($f_\mathrm{halo}=1$, $f_\mathrm{sub}=0$) of mass $10^{10} \leq M (M_\odot)\leq 10^{15}$, with a relative mass distribution function given by (\ref{haloMFeq}). The halos are assumed to be spatially distributed randomly with respect to each other. 

\item {\bf C}. Same as B, except that a fraction $f_\mathrm{sub}$ of the matter in each halo of mass $M$ is redistributed in the form of NFW subhalos in the mass range $M\times 10^{-4} \leq M_\mathrm{sub}\leq M\times f_\mathrm{sub}$, following relations (\ref{subhalofraceq}), (\ref{subhaloMFeq}) and (\ref{subhalodisteq}). 

\item {\bf D}. Same as C, except that the parent halos are assumed to follow the halo mass function (\ref{haloMFeq}) in an absolute sense, so that only a fraction $f_\mathrm{halo}(z)<1$ is in the form of halos at each redshift, whereas the remaining fraction at each redshift is assumed to be smoothly distributed $f_\mathrm{smooth}(z)=1-f_\mathrm{halo}(z)$. With a halo mass range of $10^{10} \leq M (M_\odot)\leq 10^{15}$, $f_\mathrm{smooth}(z)$ increases from 0.55 at $z\approx 0.2$ to 0.96 at $z\approx 5$.

\item {\bf E}. Same as D, except that the parent halos obey the mass function given by halo catalogues from the GIF N-body simulation (Kauffmann et al. \cite{Kauffmann et al.}). Because of resolution effects, this limits the mass range to $10^{11} \leq M (M_\odot)\leq 10^{14}$. In this case, $f_\mathrm{smooth}(z)$ increases from 0.57 at $z\approx 0.2$ to 0.97 at $z\approx 5$.

\item {\bf F}. Same as D, except that the parent halos both obey the mass function and the spatial halo distribution given by halo catalogues from the GIF N-body simulation.
\end{itemize}

The scenarios A--F are schematically illustrated in Fig.~\ref{matterdistfig}. Some of these levels of refinements correspond to assumptions adopted in the previous lensing literature. Scenario B is similar to the matter distribution model used in the lensing simulations by Bergstr\"om et al. (\cite{Bergström et al.}) for the magnification distribution of supernovae type Ia, whereas D is similar to the scenarios adopted by Oguri (\cite{Oguri}) to study the image separation distribution in macrolensed sources and by Chen et al. (\cite{Chen et al.}) to study the optical depth of small-mass halos along the line of sight to multiply-imaged quasars. 

\begin{figure}[t]
\resizebox{\hsize}{!}{\includegraphics{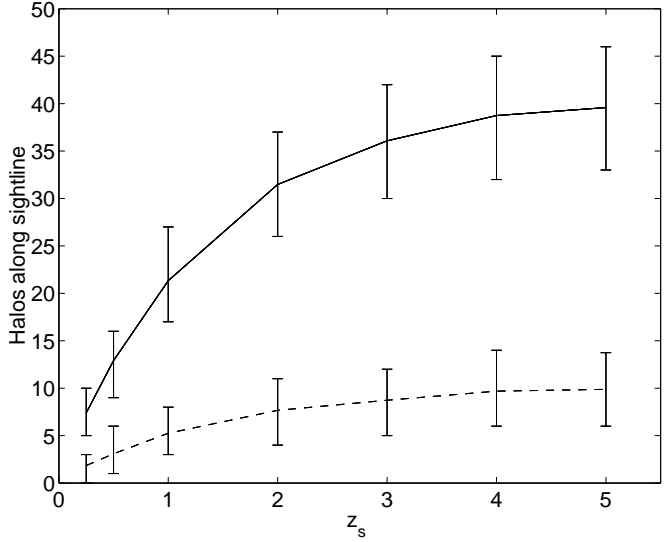}}
\caption[]{The average number of halos (solid) and subhalos (dashed) along random sightlines to sources at redshifts $z_\mathrm{s}$ in mass distribution scenario D. The error bars indicate the 68\% probability limits on the sightline-to-sightline scatter around the average numbers of halos and subhalos, respectively.}
\label{halocountfig}
\end{figure}

The matter distribution adopted by Wyithe \& Turner (\cite{Wyithe & Turner a}) in their investigation of the shortcomings of the Press \& Gunn approximation (scenario A) when estimating the distribution of microlensing optical depths, does not really correspond to any of the scenarios B--F, as they assume that no more than a single dark halo is located along each line of sight. We argue that this approach has severe limitations. In Fig.~\ref{halocountfig}, we plot the average number of halos and subhalos (with scatter indicated by errorbars) located along the line of sight in scenario D, as derived from 1000 sightline simulations. 

\begin{figure}[t]
\resizebox{\hsize}{!}{\includegraphics{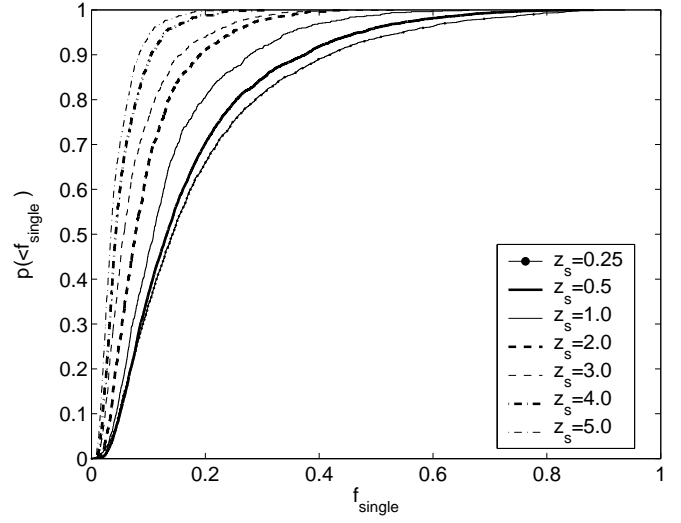}}
\caption[]{Cumulative probability distribution $p(<f_\mathrm{single})$ for the relative surface mass density contribution $f_\mathrm{single}$ from the single most dominant halo along a random sightline in matter distribution scenario D. The different lines correspond to source redshifts of $z_\mathrm{s}=0.25$ (thin solid with dots), 0.5 (thick solid), 1.0 (thin solid), 2.0 (thick dashed), 3.0 (thin dashed), 4.0 (thick dash-dotted) and 5.0 (thin dash-dotted).}
\label{surfmassprobfig}
\end{figure}

Even for sources at redshifts as low as $z_\mathrm{s}=0.25$, there are typically $\approx 7$ halos and $\approx 2$ subhalos along each sightline. As the source redshift is increased, these numbers grow monotonically, reaching $\approx 40$ for halos and $\approx 10$ for subhalos at $z_\mathrm{s}=5.0$. It should be noted, however, that these estimates include every halo for which the sightline passes inside $r_\mathrm{vir}$. As the sightline will traverse most of these halos at large distances from the centre, where the density is low, the majority of halos will actually give a very small contribution to the surface mass density of the sightline. When sightlines do happen to pass through the densest inner regions of a halo, the contribution from that object to the surface mass density can on the other hand become considerable. Hence, it is in principle still possible that the surface mass density of entire sightlines may be dominated by individual halos. To investigate how common this is, we isolate the single most dominant halo along each sightline and calculate its relative contribution to the overall surface mass density. The resulting probability distribution for the surface mass contribution of the dominant halos is plotted in Fig.~\ref{surfmassprobfig} for scenario D. The resulting probabilities for single-halo dominance are small even for nearby light sources, and drop rapidly with increasing source redshift. At $z_\mathrm{s}=0.25$, the probability of having a single halo contributing more than 50\% to the surface mass density is about 6\%, whereas already at  $z_\mathrm{s}=1.0$ this probability has dropped below 2\%. In scenario D, these low probabilities arise partly  because of the assumption of a non-zero fraction of smoothly distributed matter along each sightline, but even in scenario C, which assumes that all matter is locked up in halos (i.e. $f_\mathrm{smooth}(z)=0$), the corresponding probabilities for single-halo dominance are 26\% at $z_\mathrm{s}=0.25$ and 5\% at $z_\mathrm{s}=1.0$. Hence, individual halos are unlikely to dominate the surface mass density of sightlines to light sources at cosmological distances.  

It must of course be emphasized that this conclusion holds only for random sightlines. In the study of macrolensed quasars -- i.e. the very rare (roughly 1 in 500) objects that happen to have a foreground halo almost perfectly aligned along the line of sight -- the situation can be quite different, as demonstrated by the investigation by Wambsganss et al. (\cite{Wambsganss et al.}), who find that 90\% of the sightlines to strongly lensed sources at $z_\mathrm{s}\approx 1$ are dominated by mass concentrations at a single redshift. 

\section{Optical depth}
The microlensing optical depth $\tau$ towards a light source at redshift $z_\mathrm{s}$ is defined as:   
\begin{equation}
\tau=\int_0^{z_\mathrm{s}} \sigma(M,z,z_\mathrm{s}) \ n(M,z) \ c \ \frac{\mathrm{d}t}{\mathrm{d}z} \ \mathrm{d}z,  
\label{optdepth}
\end{equation}
where $\sigma(M,z,z_\mathrm{s})$ is the cross section for a microlens of mass $M$ at redshift $z$, and $n(M,z)$ is the corresponding number density of such objects. Formally, the microlensing optical depth represents the average number of lenses along a random line of sight. Under the assumption that the lenses do not overlap along the sightline, it also represents the fraction of sky that is covered by regions in which a point source will be microlensed. In the limit of small $\tau$ -- as e.g. in the local Universe -- the optical depth can therefore directly be used as an estimate of the microlensing probability. At higher $\tau$, this interpretation does however break down because of overlapping microlenses. While the optical depth is difficult to directly relate to observations, it is often used for quick-and-dirty estimates of the viability and relevance of different microlensing scenarios. Here, we will use it to demonstrate the effect of different assumptions concerning the spatial distribution of MACHOs along the line of sight.

In a cosmological model with flat geometry and $\Omega_\Lambda + \Omega_\mathrm{M}=1.0$, $\mathrm{d}t/\mathrm{d}z$ is given by:
\begin{equation}
\frac{\mathrm{d}t}{\mathrm{d}z} = \frac{1}{H_0}\frac{1}{(1+z)\sqrt{\Omega_\mathrm{M}(1+z)^3+\Omega_\Lambda}}.  
\label{dtdz}
\end{equation}
If the microlenses are assumed to be isolated point-masses, the cross section for microlensing can be related to the Einstein radius $R_\mathrm{E}$ through:
\begin{equation}
\sigma(M,z,z_\mathrm{s})=\pi R_\mathrm{E}^2=\frac{4\pi G M D_\mathrm{ls} D_\mathrm{ol}}{c^2 D_\mathrm{os}},  
\label{crosssection}
\end{equation}
where $D_\mathrm{ls}$, $D_\mathrm{ol}$ and $D_\mathrm{os}$ are the angular-size distances from lens to source, observer to lens and observer to source, respectively, for a lens at redshift $z$ and a source at $z_\mathrm{s}$. Under the assumption that the MACHOs are uniformly and randomly distributed along the sightline (Press \& Gunn \cite{Press & Gunn}; scenario A), their number density $n(M,z)_\mathrm{PG}$ is simply given by:
\begin{equation}
n(M,z)_\mathrm{PG}=\frac{\Omega_\mathrm{MACHO}(1+z)^3 \rho_\mathrm{c}}{M}.
\label{MACHOnumberdens_PG}
\end{equation}

By combining (\ref{optdepth})--(\ref{MACHOnumberdens_PG}) we arrive at the optical depth $\tau_\mathrm{PG}$ expected from this scenario:
\begin{equation}
\tau_\mathrm{PG} = \frac{3H_0\Omega_\mathrm{MACHO}}{2D_\mathrm{os}}\int^{z_\mathrm{s}}_{0} \frac{(1+z)^2 D_\mathrm{ls} D_\mathrm{ol} \mathrm{d}z}{\sqrt{\Omega_\mathrm{M}(1+z)^3+\Omega_\Lambda}}. 
\label{optdepth_PG}
\end{equation}

By adopting filled-beam distances (see Kayser et al. \cite{Kayser et al.} for a general discussion and Turner et al. \cite{Turner et al.} for examples of how this assumption affects the lensing optical depth), the angular size distances are given by:
\begin{equation}
D_{xy}=\frac{c}{(1+z_y)H_0}\int_{z_x}^{z_y} \frac{dz}{\sqrt{\Omega_\mathrm{M}(1+z)^3+\Omega_\Lambda} },
\label{angsizedist}
\end{equation}
where $z_x$ and $z_y$ represent the redshifts of the points $x$ and $y$.

Because of the clustering of matter, the estimate (\ref{optdepth_PG}) does not necessarily represent the true MACHO optical depth of every single sightline. In fact, most sightlines through the large-scale CDM structure of the Universe will appear underdense compared to the cosmic average. In situations where basically just a single line of sight is probed (e.g. when monitoring light sources located behind a particular galaxy or galaxy cluster), it may therefore be very useful to have a handle on the expected probability distribution of the optical depth. 

To numerically derive this distribution, we discretise the redshift axis between observer and source and compact the relevant intervening volume into $i$ lens planes, so that the optical depth becomes:
\begin{equation}
\tau = \sum_{i} f_{\Sigma,i} \sigma(M,z_i,z_\mathrm{s}) n(M,z_i) D^\mathrm{c}_i
\label{discretetau}
\end{equation} 
where $f_{\Sigma,i}$ is given by (\ref{massfrac}).
\begin{figure}[t]
\resizebox{\hsize}{!}{\includegraphics{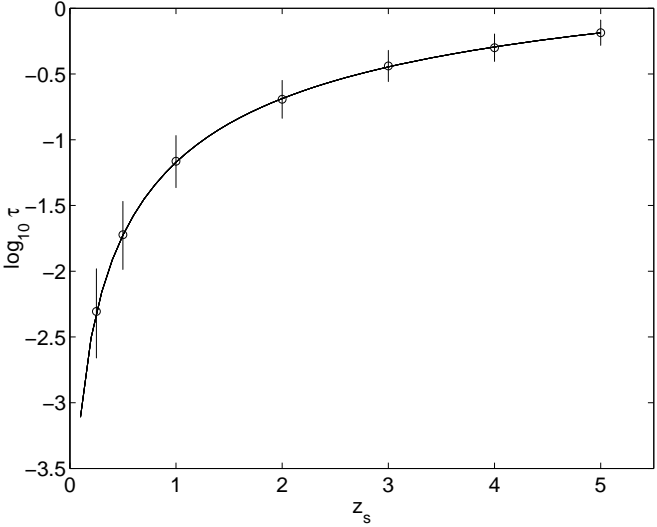}}
\caption[]{The lensing optical depth expected from the Press \& Gunn approximation (solid) compared to the distribution expected from sightline statistics for different source redshifts, using matter distribution scenario B (all matter in halos; no subhalos). The errorbars indicates the standard deviation $\sigma_\tau$ of a log-normal distribution fitted to the predicted optical depth distribution.}
\label{optdepth1}
\end{figure}

As expected, relaxing the Press \& Gunn approximation causes substantial sightline to sightline scatter in $\tau$. We demonstrate this effect in Fig.~\ref{optdepth1} by comparing the optical depth $\tau_\mathrm{PG}$ derived under the Press \& Gunn approximation to the distribution of optical depths resulting from a simulation of 1000 sightlines at different source redshifts, using (\ref{discretetau}) and matter distribution scenario B. Unless stated otherwise, the cosmological density of MACHOs is hereafter assumed to be $\Omega_\mathrm{MACHO}=\Omega_\mathrm{M}=0.3$. Since the mass-dependence of $n$ and $\sigma$ cancel in (\ref{optdepth}), the assumed mass of an individual MACHOs is of no importance here. We find that, for this scenario, the optical depth distribution at each source redshift can be reasonably well approximated by a log-normal distribution:
\begin{equation}
f(\tau,\mu,\sigma_{\ln \tau})=\frac{1}{\sqrt{ 2\pi}\tau \sigma_{\ln\tau}}\ \exp \left[-\frac{(\ln \tau-\mu)^2}{2\sigma_{\ln \tau} ^2} \right]
\label{lognormal}
\end{equation} 
with standard deviation $\sigma_{\ln \tau}$ and $\mu=\ln\tau_\mathrm{PG}-\sigma_{\ln\tau}^2/2$. As seen in Fig.~\ref{optdepth1}, $\sigma_{\ln \tau}$ decreases with increasing $z_\mathrm{s}$. The reason for this is the growing number of intervening halos at high redshifts, which reduces the relative scatter in $\tau$. While the Press \& Gunn approximation clearly improves with source redshift, the scatter remains noticeable even at $z_\mathrm{s}=5$. 

\subsection{What is the most realistic matter distribution scenario?}
\begin{figure}[t]
\resizebox{\hsize}{!}{\includegraphics{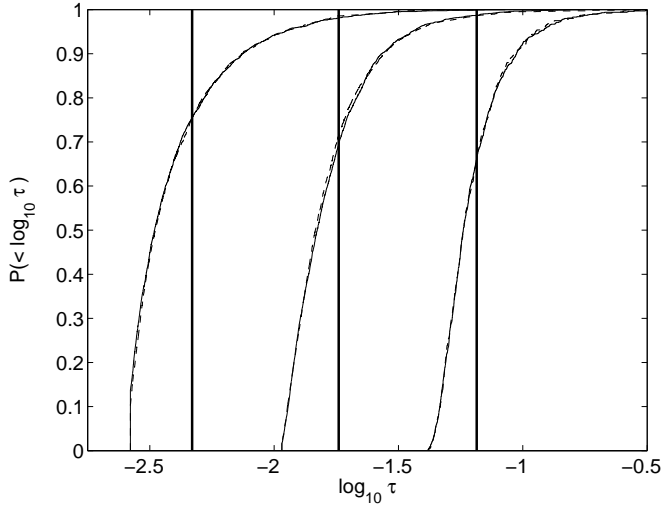}}
\caption[]{The cumulative probabilities of having a sightline with microlensing optical depth lower than $\tau$ in mass distribution scenarios E (random spatial distribution of halos; dashed) and F (spatial distribution of halos predicted by the N-body simulation; thin solid) for source redshifts of (from left to right) $z_\mathrm{s}=0.25$, 0.5 and 1.0.
The vertical, thick solid lines indicate the corresponding optical depths predicted by scenario A (randomly and uniformly distributed compact objects). The negligible difference between the predictions of scenarios E and F indicate that, at least down to the resolution limit of the N-body simulation used, the effects of halo clustering on the microlensing optical depth distribution are very small.}   
\label{clusteringfig}
\end{figure}

The predicted optical depth distribution at a given source redshift is sensitive to the details of the matter distribution scenario assumed. While option F may seem like the most sophisticated approach out of the six scenarios described in Sect. 2.5, our current implementation of this scenario is limited by the rather high halo mass resolution limit ($\sim 10^{11}\ M_\odot$) of the N-body simulation used. In Fig.~\ref{clusteringfig}, we compare the optical depth distribution predicted by scenario F to that of scenario E, in which identical assumptions concerning the mass function of dark halos are made, but in which the spatial distribution of the halos is random. As there is essentially no difference between the optical depth distributions predicted by these scenarios at any of the source redshifts investigated, we conclude that -- at least down to the resolution limit of the N-body simulation used -- the effects of halo clustering on the microlensing optical depth distribution is very small and can be neglected. The vertical part of the cumulative optical depth distributions seen at $z_\mathrm{s}=0.25$ stem from the fact that, with the limited mass resolution of the N-body simulations used, there is a non-negligible probability of having zero halos along the sightline to low-redshift sources. For these sightlines, the optical depth is instead completely dominated by the uniformly distributed matter fraction. This effect is also responsible for the artificial absence of tails towards small optical depths at higher redshifts and prevents the resulting distribution from being well-described by the lognormal fitting function (\ref{lognormal}). 

\begin{figure}[t]
\resizebox{\hsize}{!}{\includegraphics{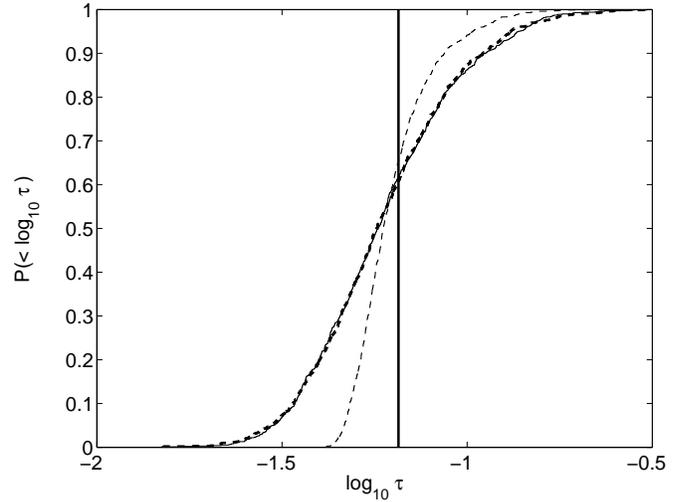}}
\caption[]{The cumulative probabilities of having a sightline with a microlensing optical depth lower than $\tau$ for a source at $z_\mathrm{s}=1.0$ in matter distribution scenarios B (all matter in halos and no subhalos; thin solid), C (all matter in halos and subhalos; thick dashed) and D (a fraction of matter in halos and subhalos; thin dashed). The vertical, thick solid lines indicate the optical depth predicted by scenario A (randomly and uniformly distributed compact objects).}   
\label{scenariocomp}
\end{figure}

In Fig.~\ref{scenariocomp}, we compare the microlensing optical depth distribution predicted by scenarios A--D for a source at redshift $z_\mathrm{s}=1.0$. As expected, the largest deviations from the Press \& Gunn prediction (A) are produced by scenarios B and C, which assume all matter to be clustered into $10^{10}$--$10^{15} \ M_\odot$ halos (i.e. $f_\mathrm{smooth}(z)=0$). These two scenarios, which differ only in the inclusion of subhalos in C, actually produce very similar optical depth predictions, indicating that subhalos represent a superfluous ingredient in high-redshift microlensing models. This result holds throughout the entire range of source redshifts investigated here ($z_\mathrm{s}=0.25$--5.0). 

Since both halo clustering (in scenario F) and subhalos (in C) seem to have a negligible impact on the results, and since the N-body simulations used have limited halo mass resolution (E), the most useful computational scheme appears to be either scenario B or D. While B formally overestimates the clustering of matter into high-mass halos, scenario D, which includes a non-zero fraction of matter distributed uniformly, involves a certain degree of artificial smoothening. Because of these computational limitations, neither scenario appears perfectly satisfying in generating realistic predictions for the line-of-sight density distributions expected in a $\Lambda$CDM Universe. In what follows, we therefore explore the consequences of {\it both} scenarios B and D, as the most realistic situation should lie somewhere in between. 

The optical depth distribution derived for our matter distribution scenarios are quite different from those presented by Wyithe \& Turner (\cite{Wyithe & Turner a}). Since most sightlines are underdense, the median $\tau$ of our distributions is always lower than $\tau_\mathrm{PG}$. The skewness of the distributions derived by us is, however, not nearly as extreme as that predicted by Wyithe \& Turner. In their model, the vast majority (90--98\%) of all sightlines have $\tau<\tau_\mathrm{PG}$, whereas in our case, the corresponding fraction is only 60--80\%. This discrepancy can be traced to their assumption of having only a single halo located along each sightline. As demonstrated in Figs.~\ref{halocountfig} and \ref{surfmassprobfig}, dark halos overlap substantially, and since their cross sections are dominated by their extended, low-density outskirts (i.e. regions with small $\tau$), the dominating effect of including more than one halo along each sightline is to reduce the probability for very small $\tau$, thereby shifting the $\tau$ distribution in the direction of the $\tau_\mathrm{PG}$ estimate.

\subsection{The predicted distribution of MACHO optical depths}
The redshift dependence of the log-normal standard deviations $\sigma_{\ln \tau}$ derived from the optical depth distribution predicted by scenarios B and D are well-described by the following polynomials:
{\setlength\arraycolsep{-30pt}\begin{eqnarray}
\sigma_{\ln \tau,\mathrm{B}}(z_\mathrm{s})=-0.004015z_\mathrm{s}^5 + 0.05854z_\mathrm{s}^4 -0.3232z_\mathrm{s}^3\\ + 0.8539z_\mathrm{s}^2 -1.156z_\mathrm{s}+1.014; \nonumber
\label{sigmatauB} \end{eqnarray}}  

{\setlength\arraycolsep{-30pt}\begin{eqnarray}
\sigma_{\ln \tau,\mathrm{D}}(z_\mathrm{s})=0.0004199z_\mathrm{s}^5 - 0.0009095z_\mathrm{s}^4 - 0.02956z_\mathrm{s}^3\\ + 0.1833z_\mathrm{s}^2 - 0.4042z_\mathrm{s}+0.4993.\nonumber
\label{sigmatauD} \end{eqnarray}}  

Using the fitting functions (23) and (24), the approximate optical depth distribution around the analytical estimate (\ref{optdepth_PG}) can easily be calculated, thereby allowing the viability of the Press \& Gunn approximation to be assessed in various microlensing situations involving cosmologically distributed MACHOs. Please note that the fitting functions are only valid for source redshifts in the range $z_\mathrm{s}=0.25$--5. Because of the negligible impact of subhalos, scenario C produces a fitting function that is virtually identical to that of D. Due of the limited halo mass resolution in scenarios E and F, the optical depth distributions generated by these are not very realistic, and cannot be well-reproduced by a lognormal function.

In scenario D, one standard deviation corresponds to a variation on the optical depth of $\approx 50\%$ for low-redshift ($z_\mathrm{s}=0.25$) light sources. While the scatter drops substantially with increasing source redshifts, it remains noticeable (at the $\approx 10$\% level) even at $z_\mathrm{s}=5.0$. In scenario B, the corresponding scatter is even higher. At $z_\mathrm{s}=0.25$, the standard deviation corresponds to a factor of $\approx 2.2$, and even at $z_\mathrm{s}=5.0$, the $1\sigma$ scatter is at the $\approx 25$\% level.

Because of the redshift dependence of the dark halo mass function, the fraction of matter that is within the halo mass resolution limits of scenario D decreases as a function of redshift. Whereas about half of the matter content is locked up in resolved halos at $z=0$, the corresponding fraction at $z=5$ is only a few percent (see Sect. 2.5). The difference between the matter distributions of scenarios B and D therefore becomes more pronounced with increasing $z_\mathrm{s}$. This does not, however, result in any dramatic difference in the $z_\mathrm{s}$ dependence of $\sigma_{\ln \tau}$ predicted by the two scenarios, since the relative contribution to the optical depth from matter at the highest redshifts is small. For example, a light source at $z_\mathrm{s}=5.0$ receives only 16\% of its optical depth from matter at $z>3$ and only 4\% from matter at $z>4$. 

The error on the $\tau_\mathrm{PG}$ estimate may become important when attempting to derive $\Omega_\mathrm{MACHO}$ from microlensing observations of only a few unique sightlines, as e.g. in the case of surface brightness fluctuations measured in galaxies selected from a small patch of the sky (Lewis \& Ibata \cite{Lewis & Ibata}). In the limiting case of only a single sightline, $\sigma_{\ln \tau}$ represents an uncertainty below which $\Omega_\mathrm{MACHO}$ cannot be determined without attempting to reconstruct the detailed surface mass density contribution of that particular sightline. A simple way to avoid this uncertainty is of course to select targets along widely separated, uncorrelated sightlines. The observational programmes proposed by Tadros et al. (\cite{Tadros et al.}) and Totani (\cite{Totani}) to monitor sources behind galaxy clusters, do for instance not suffer from this problem, since the low redshifts of their targeted foreground objects imply that sources will be selected over very large angles.

The error on the $\tau_\mathrm{PG}$ estimate may also affect, albeit to much smaller extent, estimates of the relative contribution from cosmologically distributed MACHOs to the microlensing variability displayed by a single sightline crossing a dominant mass condensation with moderately high microlensing optical depth. As an example, consider the case of a  light source at $z_\mathrm{s}=0.5$ located behind a foreground halo (galaxy or galaxy cluster) with an impact parameter such that the MACHO optical depth of that halo is expected to be $\tau_\mathrm{halo}=0.2\ f_\mathrm{MACHO}$, where $f_\mathrm{MACHO}$ represents the fractional contribution of MACHOs to the matter content of the halo. At $z_\mathrm{s}=0.5$, the optical depth of cosmologically distributed MACHOs is, in the Press \& Gunn scenario, $\tau_\mathrm{MACHO}\approx 0.06\ \Omega_\mathrm{MACHO}$ (neglecting magnification bias) or $\approx 0.018\ f_\mathrm{MACHO}$ (for $\Omega_\mathrm{M}=0.3$), if we assume that the MACHO mass fraction $f_\mathrm{MACHO}$ is scale-independent. Hence, the optical depth contribution from the cosmologically distributed MACHO population is expected to be one order of magnitude smaller than that of the MACHOs in the foreground object. If the total microlensing optical depth is observationally determined to be $\tau_\mathrm{obs}=0.08$ (with negligible observational error), one would conclude that $f_\mathrm{MACHO}=0.08/(0.2+0.018)\approx 0.37$. The uncertainty $\sigma_{\ln \tau}\approx 0.6$ predicted for this source redshift in scenario B does however translate into a 95\% confidence interval of $f_\mathrm{MACHO}= 0.37^{+0.02}_{-0.05}$ on this measurement. While this error is admittedly small, it will be compounded by the error on the $\tau_\mathrm{halo}$ estimate and may become relevant in future high-precision observations aimed to constrain $f_\mathrm{MACHO}$.
  
\section{Application to the long-term optical variability of quasars}
While the optical depth is a concept that is easy to understand and useful for demonstrating the effect of various mass distribution scenarios, it is somewhat difficult to directly relate to observations. One reason for this is that observable microlensing signatures tend to depend not only on the number of lenses affecting a sightline, but also on their velocities and where along the sightline the lenses are located. Another reason is that the optical depth, as defined here, represents the number of lenses along a random sightline, whereas flux-limited observations probe magnification-biased (rather than random) lines of sight. The calculations presented so far furthermore only take into account the effects of microlenses with a projected distance below one Einstein radius from the line of sight, whereas non-negligible magnification may in fact also result from the accumulated effects of lenses with larger impact parameters.

To demonstrate how the interpretation of observational data is affected by the use of the Press \& Gunn approximation, we couple the matter distribution model described in Sect. 2 to a microlensing code and revisit the long-standing issue of whether the optical variability of quasars on long time scales (years to decades) can be caused by microlensing by MACHOs in the subsolar mass range. In principle, this microlensing variability scenario has several attractive features. It provides a natural explanation for the statistical symmetry (Hawkins \cite{Hawkins02}), achromaticity (Hawkins \cite{Hawkins03}) and lack of cosmological time dilation (Hawkins \cite{Hawkins01}) in the optical light curves of quasars. In Zackrisson et al. (\cite{Zackrisson et al. a}) we did however demonstrate that, contrary to the claims by Hawkins, the optical, long-term variability of quasars at redshifts $z_\mathrm{s}<1$ cannot be attributed to microlensing alone. Some other variability mechanism -- most likely accretion disk instabilities -- must also contribute substantially. While this does not necessarily rule out the possibility that microlensing may contribute at some level, it complicates the procedure of using light curve statistics to infer the properties of any potential MACHO populations along the line of sight.

\subsection{Microlensing model}
To simulate the lensing situation relevant for the long-term optical variability of quasars, we couple our matter distribution model to the microlensing model described in Zackrisson \& Bergvall (\cite{Zackrisson & Bergvall}). In short, this microlensing code distributes compact objects in a number of lens planes between the source and observer. After each time step, the lenses are repositioned according to their individual velocities. The individual magnification contributions from each separate microlens is computed using analytical approximations and their total magnification computed using the multiplicative magnification approximation (Ostriker \& Vietri \cite{Ostriker & Vietri}). The most important model parameters are: the parameters describing the background cosmology ($\Omega_\mathrm{M}$, $\Omega_\Lambda$, $H_0$ and the homogeneity parameter $\eta$); the parameters describing the compact object population ($\Omega_\mathrm{MACHO}$, $M_\mathrm{MACHO}$ and their velocity dispersion $\sigma_\mathrm{v, MACHO}$ perpendicular to the line of sight) and parameters describing the source (source radius $R_\mathrm{S}$ and redshift $z_\mathrm{S}$), which is assumed to be a disk of uniform brightness. The limitations of this microlensing code are thoroughly described in Sect. 5 of Zackrisson \& Bergvall (\cite{Zackrisson & Bergvall}) and Sect. 7 of Zackrisson et al. (\cite{Zackrisson et al. a}). 

The matter distribution scenarios described in Sect. 2. are implemented in the Zackrisson \& Bergvall (\cite{Zackrisson & Bergvall}) by multiplying the density of microlenses $\rho_{i,\mathrm{ MACHO, PG}}$ in each lens plane $i$, expected in the framework of the Press \& Gunn approximation (scenario A), by the factor $f_{\Sigma,i}$ (changing stochastically from sightline to sightline) defined in (\ref{massfrac}):
\begin{equation}
\rho_{i,\mathrm{ MACHO}}=f_{\Sigma,i} \rho_{i,\mathrm{MACHO, PG}}
\label{rhoMACHO}
\end{equation}
While the lensplane-to-lensplane scatter in number of microlenses changes dramatically with this approach, we still assume a random spatial distribution of the MACHOs located in each lens plane. Hence, we assume that there is no additional small-scale clustering on scales below the resolution limit of our matter distribution model (e.g. in the form of binary MACHOs or MACHOs bound into small clusters). 

\begin{figure}[t]
\resizebox{\hsize}{!}{\includegraphics{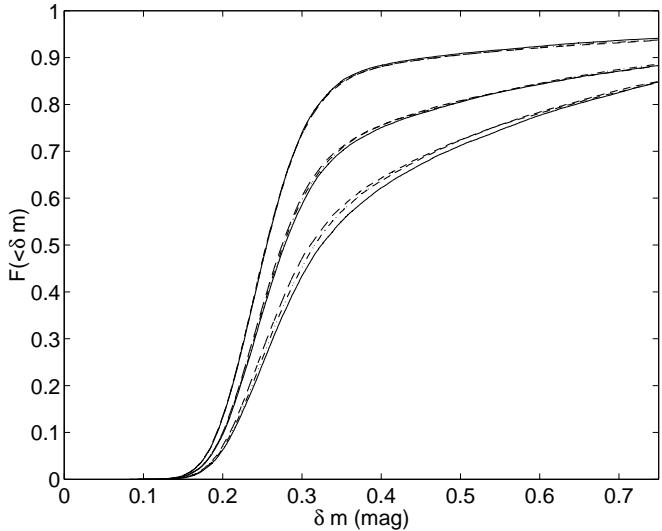}}
\caption[]{The cumulative probabilities $P(\leq\delta m)$ of observing quasars with amplitudes lower than $\delta m$ predicted in the case of $z_\mathrm{s}=0.5$, $\Omega_\mathrm{MACHO}=0.3$, $R_\mathrm{S}=10^{12}$ m and lens masses (from top to bottom at $\delta m = 0.3$) $M_\mathrm{MACHO}=10^{-2} \ M_\odot$, $10^{-3} \ M_\odot$ and $10^{-4} \ M_\odot$. For each lens mass, the different lines correspond to the results from matter distribution scenarios A (solid), B (dashed) and D (dash-dotted).}
\label{ampdistfig1}
\end{figure}

\subsection{The amplitude distribution and its redshift dependence}
To assess the contribution of MACHO microlensing to the optical long-term variability of quasars, the relevant quantity to predict is the level of brightness variability that a particular compact object population will induce during the course of an observational monitoring programme. Here, we will concentrate on the light curve amplitudes, $\delta m$, defined as the difference between the minimum and maximum of the yearly mean magnitudes $m(t)$ observed when a source is monitored for a certain period of time with a particular sampling rate:
\begin{equation}
\delta m =   \max(m(t)) - \min(m(t)).
\end{equation}
In Hawkins (\cite{Hawkins00}), this quantity was derived for several well-defined samples of quasars that had been monitored several times a year for 20 years. As demonstrated in Zackrisson et al. (\cite{Zackrisson et al. a}), microlensing predicts too small mean amplitudes at $z_\mathrm{s}<1$ and an overall amplitude distribution with too few $\delta m \geq 0.35$ mag quasars. 

To predict the amplitude statistics of microlensing-induced variability, we generate a large number of light curves with a time sampling that closely matches those of the observational samples of Hawkins (\cite{Hawkins00}). To each yearly magnification data point, derived by taking the average of four evenly spaced intrayear magnification data points, we then add Gaussian noise with standard deviation $\sigma_m=0.065$, which represents a combination of observational error and short-term variability (assumed to be intrinsic). From these yearly magnification data points, the amplitude is then derived.

As the observational samples are flux limited, we furthermore assume a quasar luminosity function and assign each light curve to a $B$-band absolute magnitude $M_B$. This quantity is converted into an apparent magnitude $m_B$ using the magnification of the first data point of the light curve and an assumed $k$-correction. All light curves that fail to meet the magnitude limit of the observational sample are then rejected. Because there is a correlation between the light curve amplitude and the magnification used to derive the apparent magnitude of each synthetic quasar (Schneider \cite{Schneider}), this procedure imposes a slight magnification bias. In the following, the $m_\mathrm{B}=21.5$ magnitude limit of the Hawkins (\cite{Hawkins00}) UVX sample will be assumed. Zackrisson \& Bergvall (\cite{Zackrisson & Bergvall}) and Zackrisson et al. (\cite{Zackrisson et al. a}) can be consulted for further details on the sample simulation procedure.  
 
As noted by Zackrisson et al. (\cite{Zackrisson et al. a}), one naively expects that relaxing the Press \& Gunn approximation would widen the distribution of light curve amplitudes, as quasars located along overdense sightlines should display increased variability, whereas quasars located along underdense sightlines should vary less. It is not obvious, however, that the effect will be very pronounced, since the amplitudes are dominated by the masses, velocities and impact parameters of the microlenses -- quantities which are unaffected by the matter distribution model explored here. What is affected is instead the probability distribution for having several microlenses simultaneously contributing substantially to a light curve.

\begin{figure}[t]
\resizebox{\hsize}{!}{\includegraphics{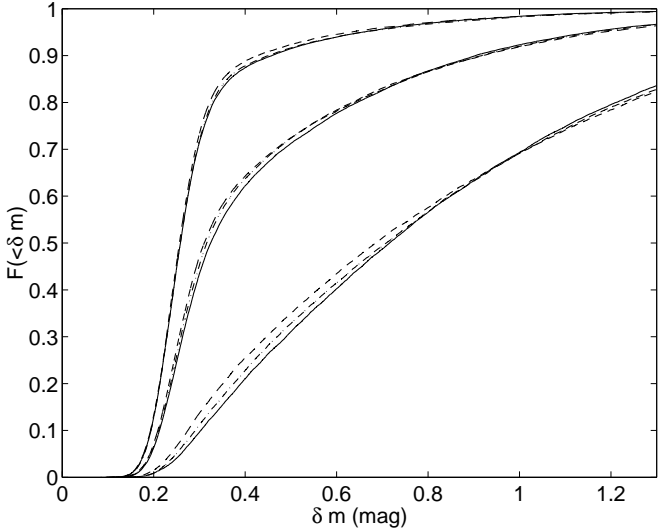}}
\caption[]{The cumulative probabilities $P(\leq\delta m)$ of observing quasars with amplitudes lower than $\delta m$ predicted for   $\Omega_\mathrm{MACHO}=0.3$, $R_\mathrm{S}=10^{12}$ m, $M=10^{-4} \ M_\odot$, $\sigma_\mathrm{v}=400$ km s$^{-1}$ in the case of matter distribution scenarios A (solid), B (dashed) and D (dash-dotted). From left to right, the three different sets of lines correspond to source redshifts of $z_\mathrm{s}=0.25$, 0.5 and 1.0.}   
\label{ampdistfig2}
\end{figure}

We find, that while relaxing the Press \& Gunn approximation does indeed affect the amplitude distribution in the way foreseen by Zackrisson et al. (\cite{Zackrisson et al. a}), the effect is typically small and varies in a non-trivial way across the relevant microlensing parameter space. The main reason for this is that when the light curve is characterized by distinct microlensing peaks, rather than multiple microlensing peaks superimposed on each other, the amplitude distribution is completely insensitive to variations in the line-of-sight density around the mean. As long as the density variations are not large enough to substantially alter the probability for peak superpositions, the Press \& Gunn approach (scenario A) therefore gives a perfectly adequate description of the amplitude distribution. When, on the other hand, the amplitudes of essentially all sightlines are dominated by overlapping microlensing peaks, density variations do have a slight effect on the amplitude probability distribution. The transition between these two regimes is illustrated in Fig.~\ref{ampdistfig1}, where we plot the amplitude distribution for a light source at $z_\mathrm{s}=0.5$ and MACHO masses in the range $10^{-4}$--$10^{-2} \ M_\odot$, under the assumption of $\Omega_\mathrm{MACHO}=0.3$ and $R_\mathrm{S}=10^{12}$ m. For a $M_\mathrm{MACHO}\geq 10^{-2} \ M_\odot$ object, the light curves typically only contain a single dominant microlensing peak, and the amplitude predictions of scenarios B and D are indistinguishable from those of scenario A. Hence, the Press \& Gunn approximation can safely be used to model the effects that the $\sim 10^{-1} \ M_\odot$ objects that may have been detected in the halos of the Milky Way and M31 (Alcock et al. \cite{Alcock et al.}; Calchi Novati et al. \cite{Calchi Novati et al.}) would have on these amplitude distributions. At $10^{-3} \ M_\odot$, a small, but noticeable difference does occur, and at $10^{-4} \ M_\odot$, the lines are clearly separated, albeit not by much. While the average amplitudes are essentially identical in the different scenarios, the amplitude distribution does become somewhat wider as a result of adopting a more realistic spatial distribution for the MACHOs. 

The source size assumed here ($R_\mathrm{S}=10^{12}$ m) is smaller than that typically used in the literature by an order of magnitude, and reflects the value found by Zackrisson et al. (\cite{Zackrisson et al. a}) to be the most promising for explaining the high level of variability observed at low redshifts ($z_\mathrm{s}<1$) in the UVX quasar sample of Hawkins (\cite{Hawkins00}). This choice allows us to assess the prospects of microlensing as an explanation for the observed variability, in light of the more realistic matter distibution scenarios discussed in the current paper. Our tests indicate that going from $R_\mathrm{S}=10^{12}$ m to the canonical value $R_\mathrm{S}=10^{13}$ m only has a modest impact on the relative difference between the amplitude distributions generated by scenarios A, B and D at the MACHO masses explored in Fig.~\ref{ampdistfig1}. 

If the microlensing parameters $\sigma_\mathrm{v, MACHO}$, $\Omega_\mathrm{MACHO}$ and $R_\mathrm{S}$ are kept fixed, the balance between single- and multiple-peak microlensing moreover depends on source redshift, as lower $z_\mathrm{s}$ correspond to a lower optical depth. This is illustrated in Fig~\ref{ampdistfig2}, where the amplitude distribution predicted in the case of a $M_\mathrm{MACHO}=10^{-4} \ M_\odot$, $\Omega_\mathrm{MACHO}=0.3$, $R_\mathrm{S}=10^{12}$ m scenario is plotted for source redshifts $z_\mathrm{s}=0.25$, 0.5 and 1.0. This corresponds to the microlensing scenario found to be one of the most promising for explaining  the long-term optical variability of quasars (Zackrisson et al. \cite{Zackrisson et al. a}; but note that no microlensing scenario was found to be completely successful in this respect). This scenario is admittedly something of a stretch, as it implies that all of the matter in the Universe (baryonic and non-baryonic) is in the form of MACHOs. This particular parameter combination has moreover been ruled out by the upper limits on the cosmological density of compact objects derived by Zackrisson \& Bergvall (\cite{Zackrisson & Bergvall}) and Zackrisson et al. (\cite{Zackrisson et al. b}). Since these investigations employed the Press \& Gunn approximation to reach these conclusions, it is nonetheless important to investigate how the conclusions from investigations are affected when this approximation is relaxed. The MACHO mass assumed here is intermediate between that advocated by Schild (\cite{Schild}), Pelt et al. (\cite{Pelt et al.}) and Colley \& Schild (\cite{Colley & Schild}) for explaining the rapid variability in Q0957+561, and the estimate advocated by Hawkins (\cite{Hawkins96}) for explaining the long-term variability of quasars that are not multiply-imaged. While certainly not favoured by the constraints on MACHOs imposed by the MACHO and EROS/EROS2 surveys (e.g. Lasserre et al. \cite{Lasserre et al.}; Tisserand et al. \cite{Tisserand et al.}), diffuse objects of this mass may still be viable, as discussed in the introduction. At $z_\mathrm{s}=0.25$, the difference between scenarios A and B/D is very small, even though the density variations along the line of sight are substantial, simply because the number of lenses affecting the light curve is not large enough for superimposed microlensing peaks to dominate the amplitude distribution. As shown in Fig.~\ref{ampdistfig2}, the difference between the predictions of the different scenarios grows slightly with increasing $z_\mathrm{s}$, as an increasing fraction of the sightlines become dominated by multiple events, even though the relative sightline density scatter goes down. 

\begin{figure}[t]
\resizebox{\hsize}{!}{\includegraphics{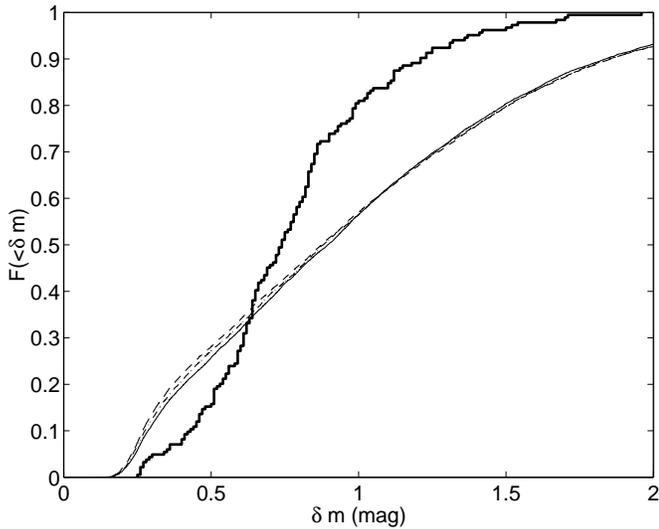}}
\caption[]{The cumulative probabilities $P(\leq\delta m)$ for amplitudes lower than $\delta m$ in the UVX sample of Hawkins (\cite{Hawkins00}; thick solid), compared to the predictions for $\Omega_\mathrm{MACHO}=0.3$, $R_\mathrm{S}=10^{12}$ m, $M=10^{-4} \ M_\odot$, $\sigma_\mathrm{v}=400$ km s$^{-1}$ and scenario A (thin solid), B (thin dashed) and D (thin dash-dotted). Here, a source redshift distribution matching that of the UVX sample has been used. While the effect of relaxing the Press \& Gunn approximation (i.e. going from scenario A to B or D) is modest, it does increase the fraction of amplitudes $\delta m\leq 0.35$ mag by a few percent, and therefore increases the discrepancy between observations and microlensing predictions.}   
\label{ampdistfig3}
\end{figure}

In Fig.~\ref{ampdistfig3} we display the amplitude distribution predicted by scenarios A (thin solid), B (thin dashed) and D (thin dash-dotted) for a distribution of quasar redshifts that matches that of the UVX sample from Hawkins (\cite{Hawkins00}), and contrast the predictions to the observed amplitude distribution of this sample (thick solid). While the effect is small, the trend is -- as suggested by  Zackrisson et al. (\cite{Zackrisson et al. a}) -- to increase the discrepancy between microlensing simulations and the observations, which is most notably seen by the lack of microlensing light curves with amplitudes $\delta m>0.35$ magnitudes. Hence, the problems associated with microlensing as the dominant mechanism for the long-term variability are in no way diminished, but instead slightly augmented, once the large-scale clustering of MACHOs is taken into account. The implementation of additional features of MACHO clustering (as described in Sect. 5) are expected to drive the amplitude distribution predicted by microlensing even further away from the observed one. Since the effect of relaxing the Press \& Gunn approximation has been shown to be quite modest (on the level of a few percent), the Press \& Gunn approach may however still be used to generate MACHO microlensing predictions of reasonable accuracy for the amplitude distribution of quasars. This greatly reduces the computational complexity of MACHO microlensing models for high-redshift sources.

Throughout this section, we have for clarity plotted the predictions of scenarios A, B and D only. As expected from the analysis of optical depth distributions in Sect. 3, the light curve amplitude predictions of scenario C are essentially identical to those of scenario B, whereas the amplitude distributions generated by scenario E are indistinguishable from those of scenario F and fall between those of scenarios A and D.

\section{Discussion}
As we have shown, the Press \& Gunn (\cite{Press & Gunn}) approximation neglects the source-to-source scatter in microlensing properties for light sources at cosmological distances. While the approximation improves with increasing source redshift, the scatter remains detectable up to a redshift of at least $z=5$, in the case where the microlenses follow the large-scale distribution of CDM. If part of the dark matter of the Universe truly is made up by MACHOs in the stellar mass range, as suggested by local microlensing detections (e.g. Alcock et al. \cite{Alcock et al.}; Calchi Novati et al.  \cite{Calchi Novati et al.}), microlensing by such objects is expected to produce observable effects in a non-negligible fraction of high-redshift light sources like supernovae type Ia and gamma-ray bursts (e.g. Minty et al. \cite{Minty et al.}; Wyithe \& Turner \cite{Wyithe & Turner b}; Baltz \& Hui \cite{Baltz & Hui}). Models of the type presented here may therefore prove to be important tools in the statistical analysis of microlensing events detected in ongoing and upcoming observations (with e.g. Swift, SNAP, JEDI or Destiny) of such objects.

The matter distribution model used here is more sophisticated than the one used by Wyithe \& Turner (\cite{Wyithe & Turner a}) in that it takes into account the combined effect of several halos along the line of sight, their spatial clustering and internal substructure. There is, however, still room for improvement, as there are a number of aspects of CDM clustering that our present model does not treat. 

The model assumes, to begin with, that all matter not locked up inside halos of mass larger than $\sim 10^{10} \ M_\odot$ (scenarios B--D) or $\sim 10^{11} \ M_\odot$ (scenarios E--F) is uniformly distributed along the line of sight. This neglects a) the spatial clustering of low-mass field halos which current cosmological N-body simulations have problems in tracing due to limited particle-resolution, and b) loosely bound matter in filaments or elsewhere that are left over by the halo finders used to create halo catalogues from N-body simulations. As this can correspond to a substantial fraction of the matter ($\gtrsim 50 \%$ in the case of the GIF simulation used here), this component should be modelled with more care. While the lower mass limit of the halo catalogues can be lowered with increasing particle resolution, taking filamentary, non-halo material into account will most likely imply using the raw N-body simulation data instead of halo catalogues and analytical density profiles, which will make the model substantially more computationally demanding. 

The current matter distribution model assumes that subhalos make up a fixed fraction of the material of each halo of a given mass, whereas the actual subhalo mass fraction displays substantial scatter (see e.g. van den Bosch et al. \cite{van den Bosch et al.}; Shaw et al. \cite{Shaw et al.}; Nurmi et al. \cite{Nurmi et al.}). As the very modest difference between the results generated by our matter distributions B and C already indicates that subhalos are unimportant for high-redshift microlensing statistics, this is not expected to have any dramatic effects on our results.    

By using the spherically averaged NFW density profiles, the effects of halo triaxiality and the associated scatter in surface mass density coming from different halo viewing angles are neglected. This situation can be improved by studying the distribution of surface mass densities produced when the line of sight traverses realistic N-body halos at a given impact parameter. If an analytical function can be fitted to this distribution, the increased scatter introduced by both triaxiality and substructure can then be statistically reproduced by Monte Carlo techniques for each NFW sphere along the sightline in the matter distribution model. We are currently investigating the feasibility of this approach (Holopainen et al. \cite{Holopainen et al.}). 

Our matter distribution model is indirectly based on results from dissipationless N-body simulations. In reality, however, CDM halos contain baryons, which are expected to alter both the shapes and density profiles of the dark halos. The general expectation is that baryon cooling inside CDM halos will cause the dark halos to contract 
(Blumenthal et al. \cite{Blumenthal et al.}; Gnedin et al. \cite{Gnedin et al.}; Sellwood \& McGaugh \cite{Sellwood & McGaugh}), although bar formation could possibly drive the evolution of the innermost part of dark halo in the opposite direction (e.g. Holley-Bockelmann et al. \cite{Holley-Bockelmann et al.}; but see Col\'in et al. \cite{Colin et al.} for a different view). Baryon cooling is furthermore expected to render dark halos more spherical (e.g. Kazantzidis et al. \cite{Kazantzidis et al.}), but the exact magnitude of this effect remains controversial.

At the current time, the velocities of the microlenses are assumed to follow a Gaussian distribution which is independent of their spatial location. In principle, this approximation can be improved by using the analytical or numerical density profile of each halo to calculate the velocity distribution of compact objects in the relevant volume of that object, and then convolve this distribution with the peculiar velocity of each halo, which may be derived from N-body simulations. For matter not associated with halos, the problems becomes substantially more complicated.  

The current model also neglects spatial clustering of microlenses on scales below the smallest structures considered (the least massive subhalos of mass $10^6 \ M_\odot$). If MACHOs are bound into binary systems or small clusters, this will affect both their position and velocity distributions in each lens plane. Without a detailed prediction for the small-scale distribution of MACHOs, it is however difficult to fully quantify the magnitude of this effect. Very dense, small-scale systems of MACHOs are of course subject to a number of dynamical constraints, as reviewed by Carr \& Sakellariadou (\cite{Carr & Sakellariadou}). 

The microlensing model used here to assess the impact of various assumptions concerning the spatial distribution of compact objects along the line of sight is based on the multiplicative magnification approximation (Ostriker \& Vietri \cite{Ostriker & Vietri}). While this is not expected to challenge the overall results of the paper, the possibility that spurious effects may arise in certain parts of the microlensing parameter space (especially in the small source regime) as a result of this, cannot be ruled out, as discussed at some length in Zackrisson \& Bergvall (\cite{Zackrisson & Bergvall}) and Zackrisson et al. (\cite{Zackrisson et al. a}). This concern can be resolved by employing ray-shooting techniques, at the expense of a substantial increase in required CPU-time. The current lensing calculations furthermore neglect contributions to the magnification bias from structures larger than the microlenses themselves (Pei \cite{Pei}). As discussed in Zackrisson \& Bergvall (\cite{Zackrisson & Bergvall}), the microlensing model used furthermore assumes filled-beam angular size distances, which is admittedly not a self-consistent treatment of the situation, and may be a bad approximation in certain parts of the microlensing parameter space. The sightline-to-sightline variations predicted by the matter distribution scenarios considered here act to augment this problem even further. To remedy this problem would however require a much more sophisticated lensing algorithm, e.g. of the Holz \& Wald (\cite{Holz & Wald}) type, in which ``each photon creates its own Universe'' along its path from source to observer. For a fixed $\Omega_\mathrm{MACHO}$, the effect of decreasing the filling factor of each beam (i.e. going from filled to empty) is to decrease the the optical depth (e.g. Turner et al \cite{Turner et al.}). Because of this, the sightlines that correspond to low microlensing optical depths in our formalism would tend to have them lowered even further if the filled-beam approximation was dropped, thereby increasing the sightline to sightline scatter. This effect is however not expected to be dramatic unless the source redshift is very high, as it is only then that the filled-beam and empty-beam angular size distances differ substantially. By evaluating equation (\ref{optdepth_PG}) when shifting between filled-beam and empty-beam angular size distances, we find that the difference in optical depth between these two extreme situations (the most realistic situation should lie somewhere in between) is $\leq 3$ \% for source redshifts of $z_\mathrm{s}\leq 1$, which is negligible compared to the scatter induced by other components of the the matter distribution model. However, at $z_\mathrm{s}=5$ the difference is $\leq 23$ \%, which is sufficiently large to warrant consideration in future, high-precision predictions of the optical depth distribution. 

\section{Summary}
When predicting the gravitational microlensing effects of dark matter in the form of compact objects (MACHOs) on high-redshift light sources, it is customary to adopt the simplifying assumption that MACHOs are randomly and uniformly distributed along the line of sight (the Press \& Gunn \cite{Press & Gunn} approximation). In this paper, we explore the consequences of relaxing this assumption in favour of a model where the microlenses follow the spatial clustering of CDM, as would be expected for scenarios involving non-baryonic MACHOs. In summary, we find that:
\begin{itemize}
\item The contribution from cosmologically distributed MACHOs to the microlensing optical depth of individual sightlines may differ from the Press \& Gunn estimate (\ref{optdepth_PG}) by a non-negligible factor, which depends on the redshift of the light source studied. Here, we quantify the scatter as a function of source redshift and demonstrate that the optical depth of an individual sightline can be incorrect by a factor of $\approx 2$ (at the $1\sigma$ level) for a light source at $z_\mathrm{s}=0.25$. While the optical depth scatter decreases with source redshift, it remains noticeable even at $z_\mathrm{s}=5$. This result implies that attempts to determine the cosmological density of MACHOs from microlensing observations of a single or only a few random sightlines will be subject to substantial uncertainties. 
\item The sightline-to-sightline scatter of microlensing optical depth is dominated by the clustering of MACHOs into galaxy-mass CDM halos. Both the spatial correlation of these halos and their internal clumpiness due to substructure is shown to have negligible impact on the optical depth statistics.
\item While relaxing the Press \& Gunn (\cite{Press & Gunn}) approximation in favour of the matter distribution scenarios explored here is shown to have a rather modest impact on the MACHO microlensing predictions for light curve amplitudes, the microlensing predictions are driven further away from the observed statistical properties of the long-term optical variability of quasars that are not multiply-imaged. Hence, the problems with microlensing as the primary cause for this variability are in no way diminished, but instead slightly augmented, once the large-scale clustering of MACHOs is taken into account. 
\end{itemize}

\begin{acknowledgements}
EZ acknowledges research grants from the Swedish Royal Academy of Sciences, the Academy of Finland and the Swedish Research Council. We thank the anonymous referee for useful comments on the manuscript and Pasi Nurmi for providing the code used to generate the dark halo mass functions.   
\end{acknowledgements}

\end{document}